**Abundant ammonium hydrosulphide embedded in cometary dust grains**

K. Altwegg[1], M. Combi[2], S. A. Fuselier[3,4], N. Hänni[1], J. De Keyser[5], A. Mahjoub[6,7], D. R. Müller[1], B. Pestoni[1], M. Rubin[1], and S. F. Wampfler[8]

[1] Physics Institute, Space Research & Planetary Sciences, University of Bern, Sidlerstrasse 5, 3012 Bern, Switzerland

e-mail: kathrin.altwegg@unibe.ch

[2] Department of Climate and Space Sciences and Engineering, University of Michigan, Ann Arbor, MI, USA

[3] Space Science Directorate, Southwest Research Institute, San Antonio, TX, USA

[4] Department of Physics and Astronomy, The University of Texas at San Antonio, San Antonio, TX, USA

[5] Royal Belgian Institute for Space Aeronomy, BIRA-IASB, Brussels, Belgium

[6] Jet Propulsion Laboratory, California Institute of Technology, Pasadena, California

[7] Space Science Institute, 4765 Walnut St, Suite B, Boulder, CO 80301

[8] Center for Space and Habitability, University of Bern, Gesellschaftsstrasse 6, 3012 Bern, Switzerland

**ABSTRACT**

Ammonium hydrosulphide has long since been postulated to exist at least in certain layers of the giant planets. Its radiation products may be the reason for the red colour seen on Jupiter. Several ammonium salts, the products of $NH_3$ and an acid, have previously been detected at comet 67P/Churyumov-Gerasimenko. The acid $H_2S$ is the fifth most abundant molecule in the coma of 67P followed by $NH_3$. In order to look for the salt $NH_4^+SH^-$, we analysed in situ measurements from the Rosetta/ROSINA Double Focusing Mass Spectrometer during the Rosetta mission. $NH_3$ and $H_2S$ appear to be independent of each other when sublimating directly from the nucleus. However, we observe a strong correlation between the two species during dust impacts, clearly pointing to the salt. We find that $NH_4^+SH^-$ is by far the most abundant salt, more abundant in the dust impacts than even water. We also find all previously detected ammonium salts and for the first time ammonium fluoride. The amount of ammonia and acids balance each other, confirming that ammonia is mostly in the form of salt embedded into dust grains. Allotropes $S_2$ and $S_3$ are strongly enhanced in the impacts, while $H_2S_2$ and its fragment $HS_2$ are not detected, which is most probably the result of radiolysis of $NH_4^+SH^-$. This makes a prestellar origin of the salt likely. Our findings may explain the apparent depletion of nitrogen in comets and maybe help to solve the riddle of the missing sulphur in star forming regions.

## 1 INTRODUCTION

Ammonium salts in the interstellar context are poorly understood. They have mostly sublimation temperatures above water and are therefore in a semi-refractory / semi-volatile state on grains, hard to detect by remote sensing. Semi-refractory / semi-volatile here means that the species sublimate at temperatures ≥ the sublimation temperature of water, but < normal refractories like e.g. minerals. They are volatile enough, so that they can be detected in the gas phase at elevated temperatures, but disappear from warm grains in a relatively short time and cannot be detected in dust grains brought back by e.g. the Stardust mission or by the dust mass spectrometer COSIMA on board Rosetta (Kissel et al.2007), because they sublimate before they can be analysed. E.g. $NH_4Cl$ sublimates at 473 K (Clementi E. & Gayles J. N. 1967). Once they sublimate, they dissociate into ammonia and acid, almost indistinguishable from the pure molecules from ices. Meanwhile, however, it has been established that the 6.85 µm ice band seen near young stellar objects may well be due to $NH_4^+$ (Schutte & Khanna 2003). In interstellar ice, a feature at 4.62 µm is identified as $OCN^-$, a possible anion (e.g. Pendleton et al. 1999). However, so far these are the only signs of ammonium salts outside the Solar System.

Clear signatures of ammonium salts in a cometary dust outburst have been previously detected by the ROSINA mass spectrometer (Balsiger et al. 2007) on the Rosetta mission around 67P/Churyumov-Gerasimenko (Altwegg et al. 2020). Five different salts have been identified by their sublimation products in the cometary coma: $NH_4^+CN^-$, $NH_4^+OCN^-$, $NH_4^+Cl^-$, $NH_4^+HCOO^-$, $NH_4^+CH_3COO^-$. The infrared spectrometer (VIRTIS) on board Rosetta detected a signature around 3.2 µm from spectra of the nucleus surface, which is also compatible with ammonium salts (Poch et al. 2020). This signature seems to be ubiquitous across the surface of the comet. Ammonium formate, ammonium sulphate, or ammonium citrate match VIRTIS reflectance spectra best. The feature around 3.2 µm is known to be part of the signature of $NH_4^+$ together with the more prominent feature at 6.85 µm (Schutte et al. 2003), not measured by Rosetta. This is valid for many ammonium salts and the exact wavelength of the 3.2 µm band depends on several parameters, not only the associated anion, but also e.g. on temperature or ice matrix. The anions mostly evade detection using remote sensing. In the Solar System, solid ammonium hydrosulphide ($NH_4^+SH^-$) has long since been postulated to be a key ingredient in some layers of the giant planets (Weidenschilling & Lewis, 1973; Atreya et al. 1999; Roman, Banfield & Gierasch 2013). Its radiation products may be responsible of the color of Jupiter's great red spot (Loeffler et al. 2016). Ammoniated minerals, most probably saponite, have been identified on Ceres and some other asteroids (King et al. 1992; Berg et al. 2016).

Ammonium salts are the products of ammonia and almost any acid. They form readily at low temperatures (Altwegg et al. 2020 and references therein, Table 1 in supplementary material). It is therefore not surprising that such salts are found on comets, where it is known that most nitrogen is in the form of $NH_3$ ((0.67 ± 0.20) % rel. to water (Rubin et al. 2019)). Looking at possible acids, $H_2S$ with (1.10 ± 0.46) % relative to water (Rubin et al. 2019) is the fifth most abundant molecule in the bulk ice of 67P after $H_2O$, $CO_2$, CO and $O_2$ and by far the most abundant acid. Therefore, it makes sense to look for the presence of $NH_4^+SH^-$. This, however, is not simple in the coma as $NH_3$ and $H_2S$ are both individual parts of cometary ice and that upon sublimation the ammonium salts quickly dissociate into $NH_3$ and the acid again. $NH_4^+SH^-$ is unlike many other salts found on Earth in that it is unstable under typical atmospheric conditions and room temperature. In addition, its sublimation products are not desired in laboratories due to their toxic and corrosive nature. Therefore, laboratory data are rather scarce for this salt.

While the ROSINA instrument was built for the analysis of neutral and ionized gases in the coma of 67P, it turned out that it could also detect at least one group of grains. These are grains which contain volatile and semi-volatile components (Pestoni et al. 2021a, 2021b) sublimating in or near the ion source of the instrument. The detection of ammonium salts with ROSINA on Sep. 5, 2016 was related to such a dust impact (Altwegg et al. 2020). The signature of $NH_4^+$ seen by VIRTIS is from the surface of the nucleus, which also consists mainly of dust. Ammonium salts, therefore, are most likely part of cometary grains. In order to detect the salts, we have to look in the spectra from the ROSINA mass spectrometer for signs of dust impacts, which produce sudden density increases of $NH_3$ and possible acids above the "normal" background coma values. To strengthen the case, for some ammonium salts, there are also small signatures of byproducts of the sublimation of salts besides the pure acid (Hänni et al. 2019), for example formamide is produced by the sublimation of $NH_4^+HCOO^-$. For $NH_4^+SH^-$, laboratory experiments have found some $NS_x$ and $S_3$ after irradiation by energetic protons (Loeffler et al. 2015), which may also be indicative of the salt.

Here we report the first detection of $NH_4^+SH^-$ and, in addition, $NH_4^+F^-$ in the coma of 67P. In all cases, the appearance of ammonium salts is indeed related to dust.

## 2 INSTRUMENTATION

The Rosetta spacecraft reached comet 67P/Churyumov-Gerasimenko around Aug. 1, 2014, by which time it had matched the velocity of the comet. For more than two years, it orbited the comet at different cometocentric distances, travelled from 3.6 au through perihelion at 1.24 au, and out again to 4 au, before it softly crash-landed on the comet on Sep. 30, 2016. Included in the payload was the ROSINA (Rosetta Orbiter Sensors for Ion and Neutral analysis) suite of instruments with two mass spectrometers and a pressure sensor (Balsiger et al. 2007). In this paper, we mostly use data from the high-resolution Double Focusing Mass Spectrometer (DFMS). To account for variability of the background coma density, we also use data from the nude gauge of the COmetary Pressure Sensor (COPS), which measures the total density of the coma.

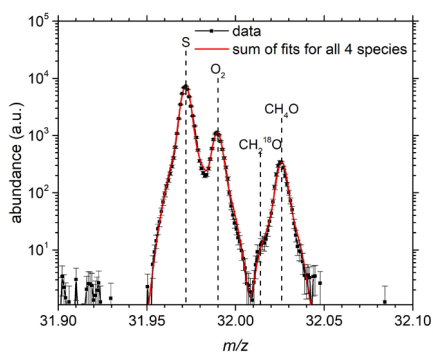

Figure 1: Mass spectra taken on January 11, 2016, 83 km from the nucleus. The red line shows the sum of individual Gaussian fits for the 4 species on $m/z$ 32, where their exact masses are indicated by dashed lines.

DFMS is a classical magnetic mass spectrometer in Nier-Johnson configuration with a field of view of 20°. The mass resolution $m/\delta m$ is 9000 FWHM at $m/z$ 28 (3000 at the 1% level). Neutral gas is ionized by electron impact, which not only yields the parent molecule, but also some fragments. The detector consists of a multichannel plate (MCP) in analogue mode with two redundant anodes, each of which has 512 pixels. Information on the instrument and data analysis are in Balsiger et al. (2007), LeRoy et al. (2015) and, especially for sulphur bearing species, in Calmonte et al. (2016). Fig. 1 shows a mass spectrum of $m/z$ 32, which demonstrate the mass resolution and the dynamic range of the instrument. The data analysis is a two-step process. Because the MCP detector works in analogue mode, the detector signal (charge deposited on the anode during the integration interval) depends not only on the number of ions impacting the detector, but on their energy, on their molecular structure and size and on the gain of the detector, which is adjusted to the signal strength for each spectrum. In the first step, we take into account the energy dependence and the gain (see e.g. De Keyser et al. 2019), but not the structure of the species. The y-axis labelled abundance in arbitrary units, therefore, corresponds only approximately to number of ions, as we do not consider the different structures of the species. The peaks can easily be fitted by two Gaussians (De Keyser et al. 2019), which then yields the total abundance per contributing species. It is clear from Fig. 1 that relative statistical errors are very low in these cases and can be ignored compared to other uncertainties.

To go from abundances measured on the detector to densities in the ion source, we need the ionization cross section, the fragmentation pattern, the ionizing electron current (fixed at 200 µA), the detector sensitivity and the instrument transfer function for all species. The ionization cross section and the detector sensitivity have to be calibrated for all species, especially because DFMS uses 45 eV electrons, while most of the literature contains data for 70 eV. This cannot be done for many of the ammonium salts, as their products are corrosive and/or poisonous. Therefore, we assume here an ionization cross section equal 1 for all species and do not give absolute densities. We analyse for all species M the parent ion $M^+$ and then correct for fragments on different masses according to the NIST database (Steins S. E., 2018). In principle, we could also use the measured daughter abundances. However, contrary to the parent ions $M^+$, daughter species often have contributions from different parents, which makes this difficult. E.g. for $H_2S$, the relative intensities according to NIST are 100 ($H_2S^+$), 42 ($HS^+$) and 44 ($S^+$), which means that the peak on $m/z$ 34 corresponds to ~ 55% of the total ionized $H_2S$. Measured $S^+$ likely also has an unknown contribution from e.g. $S_2$ and other sulphur bearing molecules. By using NIST data however, we introduce another uncertainty due to the difference in electron energy.

While DFMS operates in many different measurement modes, most of the time it stepped through integer masses one by one, always starting and ending with $m/z$ 18. The original mass range $m/z$ 13 to $m/z$ 100 was extended during the mission to $m/z$ 140 and later to $m/z$ 180. Integration time per mass was 20s. Due to adjustments of the detector gain, the overhead time per mass was up to 10s. This means, for example, that the time difference between measurements of $m/z$ 17 (OH, $NH_3$) and $m/z$ 32 (S, $O_2$, $CH_3OH$) was approximately 7 min.

## 3 MEASUREMENTS
### 3.1 Overall mission data

Fig. 2, panel a) shows all usable spectra of the $H_2S/NH_3$ abundance ratio over the mission together with the heliocentric distance. The abundances in Figure 2 are interpolated to the same measurement time wherever possible. A perfect correlation should yield a straight horizontal line. From this plot, there is no obvious correlation between the two species. The small and large modulations are mostly due to different latitudinal dependence, where the sublimation of the two species is different for the northern and southern hemisphere, depending on the changing subsolar point and S/C position during the mission. This may be a result of different volatilities (sublimation temperature $T_{sub}$ ~ 85K and 105K for $H_2S$ and $NH_3$ respectively (Minissale et al. 2022)). Fig. 2 shows a large spread of $NH_3/H_2S$ ratios over the mission from 0.1 inbound around 3 au to >10 outbound at 3.8 au. A similar picture can be deduced from Fig. 3. This lack of correlation means that the cometary ice contains $NH_3$ and $H_2S$, but predominantly not in the form of the salt $NH_4^+SH^-$.

The measured local abundance depends on several parameters like the cometocentric distance, the S/C latitude and longitude with respect to the nucleus and the S/C attitude. Because of the time difference between measured $NH_3$ and $H_2S$ abundances it makes sense to normalize them which eliminates sudden changes in the background coma density affecting all nucleus gases simultaneously. The main contribution to S is from $H_2S$. The abundance of $O_2$ follows the water density quite closely over the whole mission (Läuter et al. 2020). OH is clearly a fragment of water. Therefore, it is valid to normalize $NH_3$ to OH (*m/z* 17) and S to $O_2$ (*m/z* 32) as is done in Figure 2. The normalization cancels the dependence on cometocentric distance if we assume a $1/r^2$ Haser model.

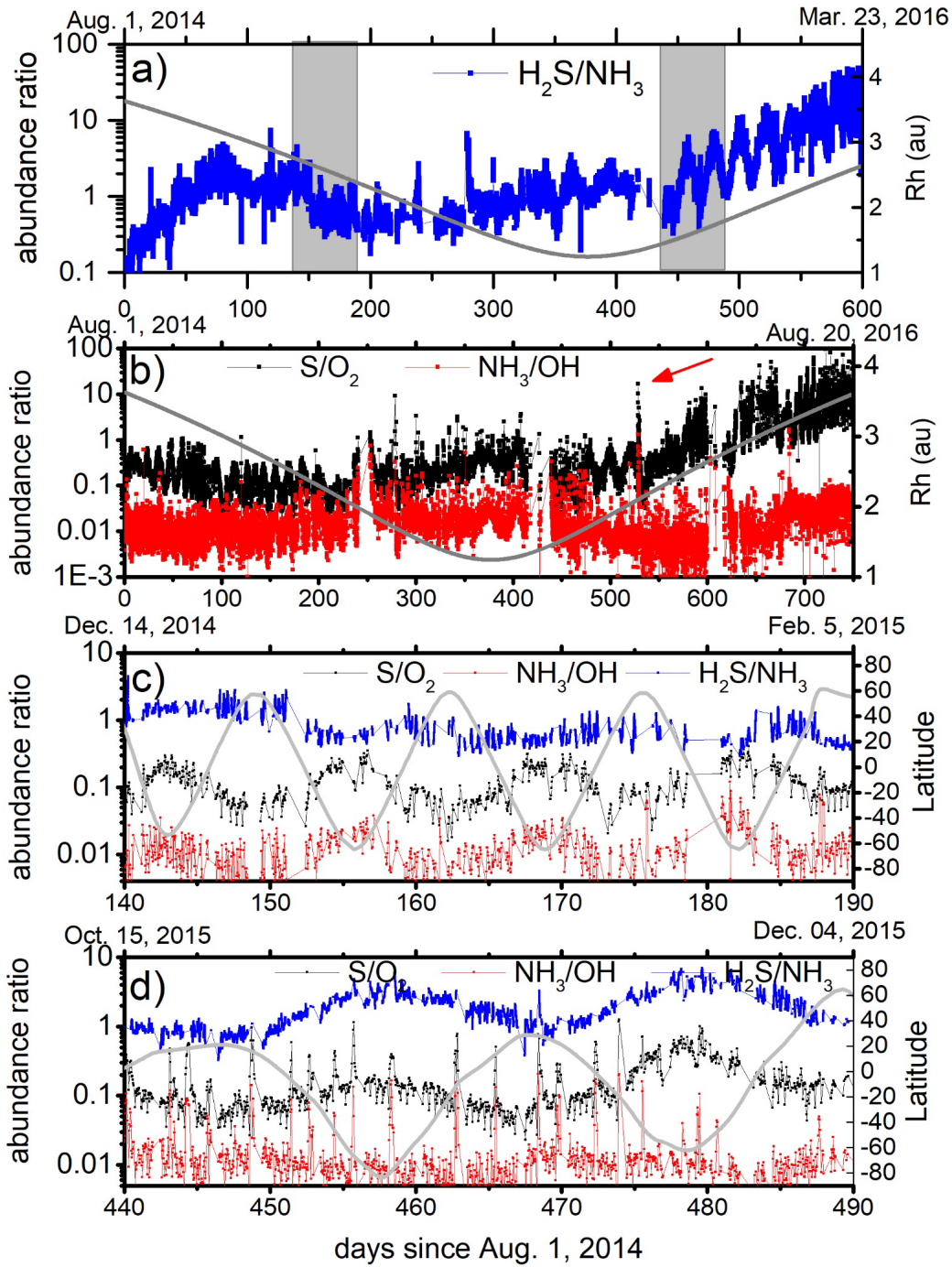

Figure 2: Ratios of $H_2S/NH_3$ (blue), $NH_3/OH$ (red) and $S/O_2$ (black) abundances for different periods of the mission. In the panels a) and b), the heliocentric distance Rh (dark grey) is also shown. In panels c) and d), the sub-S/C latitude (degrees) is shown (light grey). The grey bars in panel b) indicate the periods for the two zooms. The red arrow indicates the event of January 11, 2016.

There are no differences in detector gain between the two species on the same mass as they are measured simultaneously. What remains are latitudinal effects as $NH_3$, $H_2S$, $H_2O$ and $O_2$ sublimate differently from the northern and southern hemispheres (Läuter et al. 2020) and also effects of heliocentric distance. However, these are gradual changes compared to dust impacts. Fig 2, panel b) shows the ratios of $NH_3/OH$ and $S/O_2$ over almost the complete mission. Panel c) is a zoom in for a period outside of 2 au pre-perihelion and panel d) is a zoom for a period inside of 2 au post-perihelion. Panels c) and d) also contain $H_2S/NH_3$ for the periods, when interpolation was possible (i.e., no fast slewing).

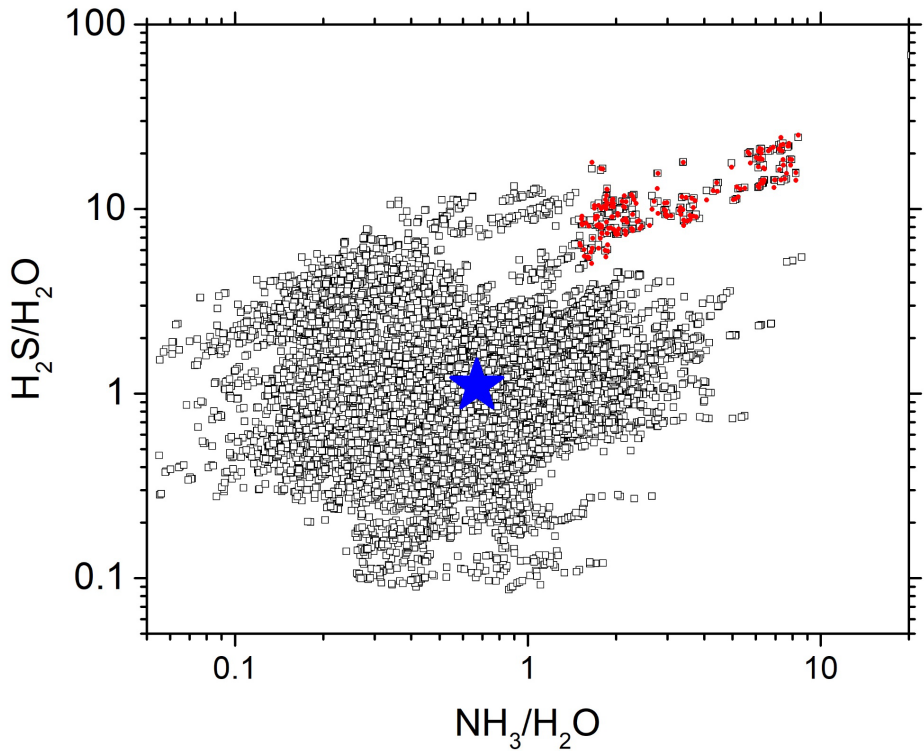

Figure 3: Relative abundances compared to water in %. The abundances are interpolated to the same measurement time wherever possible. Black squares: all measurements over the mission, red dots: measurements associated with impacts, blue star: bulk value (Rubin et al. 2019).

From panel b), it is evident that the ratio of $NH_3/OH$ is relatively constant over the mission, certainly much more than $H_2S/NH_3$. There is a slight enhancement during the perihelion passage and towards the end of the mission, although the latitudinal dependence is different for the two species (panels c) and d)). For $S/O_2$, there are relatively strong latitudinal effects (panels c) and d)) and a strong increase of the ratio for the last part of the mission. This increase is due to the decrease of $O_2$ with heliocentric distance, which was steeper than for $H_2S$ post-perihelion (Läuter et al. 2020). Also seen in panel b) is a strong scatter in the ratios inside of ~2.25 au, not seen for $H_2S/NH_3$. Zooming into this period (panel d)), it becomes clear that this is not an arbitrary scatter. Rather, the ratios of $S/O_2$ and $NH_3/OH$ show random, but simultaneous strong narrow peaks, lasting for three to seven measurement points. This periodicity correlates to 2-5h. The strongest peak appeared on day 528 (red arrow in the top panel), which corresponds to Jan. 11, 2016. These peaks are the signature of dust impacts, which contain clearly abundant $NH_3$ and $H_2S$. In these peaks, $NH_3$ and $H_2S$ show a very good correlation, seen also in Fig. 3, which shows the relative abundances of $H_2S$ vs. $NH_3$ for the entire mission. The Pearson correlation R for the entire data set is low (0.19, ~32000 data points), while R is relatively high (0.81, 204 data points) for the measurements which we associate with dust impacts (red dots).

$NH_3$ and $H_2S$ peak simultaneously for ~60 dust impact events over the mission. Figure 4 is a histogram plot of these events as a function of month, together with the heliocentric and cometocentric distance. This plot must be considered with some caution, as there were months where DFMS was not always on (e.g. Sept. 22 to Oct. 14, 2015) due to very large distances to the comet or where the spacecraft (S/C) was in safe-mode (e.g. March 30 to April 11, 2015). This might add another ~5 / 10 events to the Sept./Oct. periods, respectively and ~2 to the April period. Thus, the number of events is not normalized by the total operation time. However, it is evident that all events occurred inside of 2.25 au and almost all of them when the southern hemisphere was in summer and the number of events is higher post-perihelion. The event closest to the nucleus occurred when the spacecraft was at a distance of 83 km (Jan. 11, 2016). Grains travel with a velocity of 0.3-15 ms$^{-1}$ (Della Corte V. et al. 2016), which means they need between 2h and up to 2 days to reach the S/C. With a rotation period of the comet of ~12 h, it is impossible to derive the foot point (latitude, longitude) on the nucleus where these grains originate. However, it is plausible that they mostly come from the southern hemisphere, which was much more active post-perihelion.

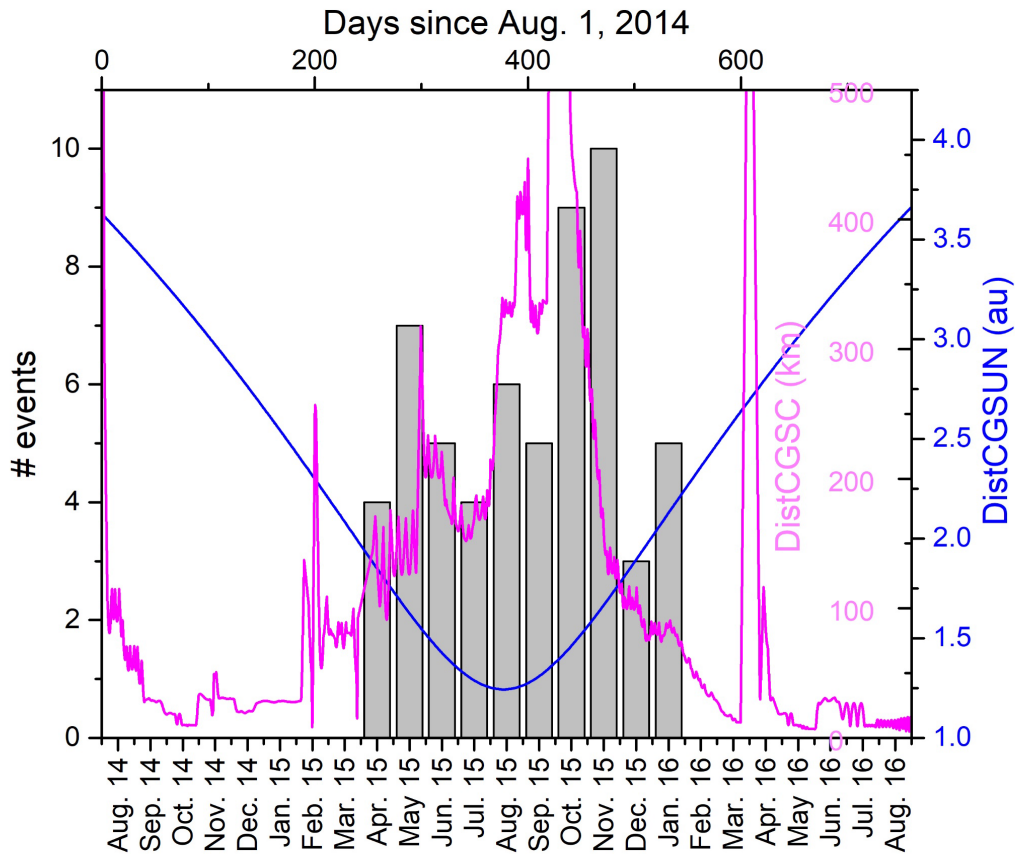

Figure 4: Number of events with sudden, high $NH_3$ & $H_2S$ abundances (more than a factor 5 above the background coma) per month together with the heliocentric distance (blue) in au and the distance between S/C and comet (magenta) in km.

Looking at the details of single events, it is apparent that boundary conditions of their occurrence vary quite considerably. There are events, which happened when the S/C and the instrument were in steady measurement mode, no slewing, and no mode changes. For other events, the S/C was actually slewing, or the event happened during the warm-up phase of the instrument shortly after or even during a wheel-off loading or a manoeuvre, when the instrument was off. The height of the peaks also varies, which may mean that some grains entered directly into the ion source of DFMS while others ended up close to the entrance of DFMS. These complications make it impossible to derive meaningful absolute quantitative abundances for the different species. However, we can derive the undisturbed coma background from spectra measured as close to the peaks or at least with very similar boundary conditions (longitude, latitude, switch-on time, off-nadir angle) in the same period. Comparing these spectra with the spectra in the peaks provides insight on the composition of semi-volatile molecules in / on these grains.

Table 1: Boundary conditions for four typical events analysed in this paper.

| Event # | Date / time ($m/z$ 17) | Helio-centric dist. (au) | Com-etocen-tric dist. (km) | Off-nadir angle (degree) /slew | Ion source temperat-ure (°C) | Background pressure (mbar) for the impact time and the comparison spectrum | Time since last switch-on | Date /time of back-ground spectrum ($m/z$ 17) |
|---|---|---|---|---|---|---|---|---|
| 1 | Apr. 12, 2015 0637 h | 1.88 | 145 | 1.7 | -20 to -10 | 2.6e-10 / 3.7e-10 | >2 h | Apr. 15, 2015 2055 h |
| 2 | May 27, 2015 2249 h | 1.55 | 308 | 2.44 | 80 to 60 | 3.3e-10 / 2.5e-10 | 7 min | May 28, 2015 1828 h |
| 3 | Nov. 12, 2015 0930 h | 1.65 | 175 | Slew | 30, stable | 6e-10 / 1.8e-9 | >2 h | Nov. 12, 2015 0730h |
| 4 | Jan. 11, 2016 1512 h | 2.1 | 81.5 | 4.3 | 6, stable | 3.3e-10 / 3.5e-10 | >2 h | Jan. 11, 2016 1430h |

### 3.2 Results for four typical events

In this paper, we concentrate on four typical events (Table 1) with diverse boundary conditions. This is done on purpose to show that the results are very similar, independent of heliocentric distance, distance to the comet and status of the instrument (e.g. ion source temperature). In all four cases, COPS did not register any dust impacts, indicating a steady background coma from sublimating nucleus ices while DFMS registered quasi-simultaneous peaks (at least five times larger than the coma background) for $H_2S$ and $NH_3$ (see Fig. 2, bottom right panel). We corrected all background abundances with the pressure measured by the COPS nude gauge during the two spectra (background and peak) as the measured background coma varied, especially for the case of the slew. The cleanest event is #4, which is the peak observed on Jan. 11, 2016. This event was a clear impact into the ion source of DFMS. It was not as violent as the one on Sep. 5, 2016, described in Altwegg et al. (2020), where the S/C environment was full of dust and where the impact generated some instrument anomaly during the measurement of the mass range $m/z$ 32-34, exactly where we expect $H_2S$. The Jan. 11, 2016 event did not cause any anomaly and happened between $m/z$ 97 and $m/z$ 98.

Before this event, the coma was quiet and the spectra of high quality. The instrument was in steady state, including the temperature in the ion source. Fig. 5 shows the spectra around the time when the impact happened.

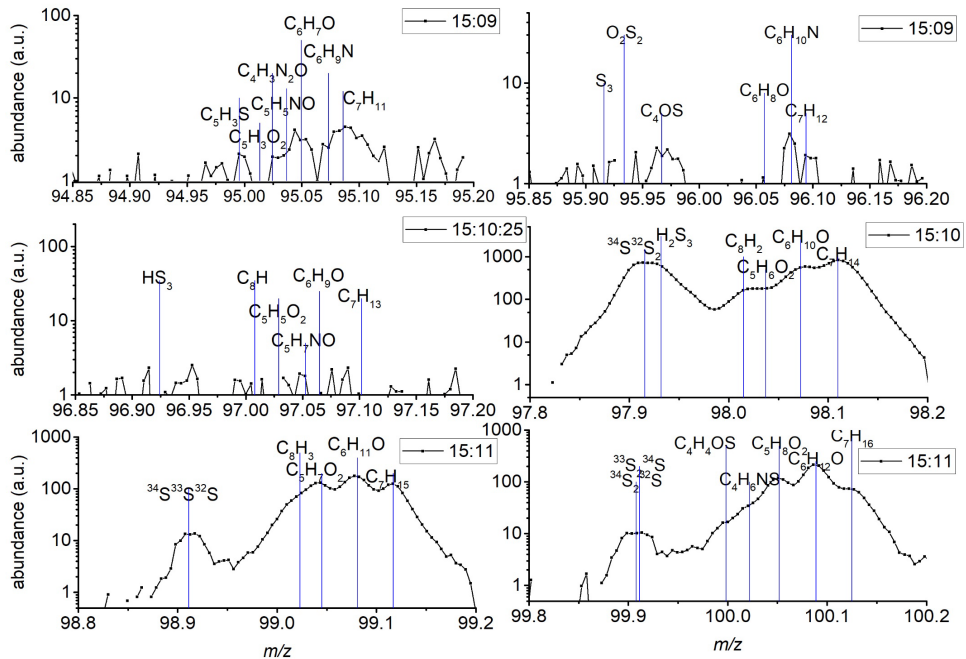

Figure 5: Mass spectra (abundance in arbitrary units) from *m/z* 95 to *m/z* 97 before and from *m/z* 98 to *m/z* 100 after the impact on Jan. 11, 2016 around 15:10h UTC.

Between *m/z* 95 and *m/z* 97, the signal was mostly noise with no identifiable peaks. In the plot, we indicate the position of possible species, but the signal to noise ratio does not allow positive identification for any of them. On *m/z* 98 and following, peaks were clearly identifiable with quite large amplitudes. While there is e.g. no peak at the position of $^{32}S_3$ on *m/z* 96, which is before the impact, there are clear peaks for the minor isotopologues ($^{32}S_2^{34}S$, $^{32}S^{33}S^{34}S$ and $^{32}S^{34}S_2$) after the impact. The density in the ion source, however, is not stable enough at this time to derive any isotopic ratios.

Fig. 6 shows a comparison of masses relevant for ammonium salts. In black and grey are spectra after the impact, in red spectra immediately before the impact. The differences are obvious. $NH_3$ increased by a factor of ~660, S and $H_2S$ by more than a factor 100, while water increased by about a factor 3 and $O_2$ decreased slightly. Other highly volatiles like e.g. $CH_4$ (not shown here) also increased similar to water. FH was not seen before the impact with an upper limit of 1. HCl increased by a factor of 160.

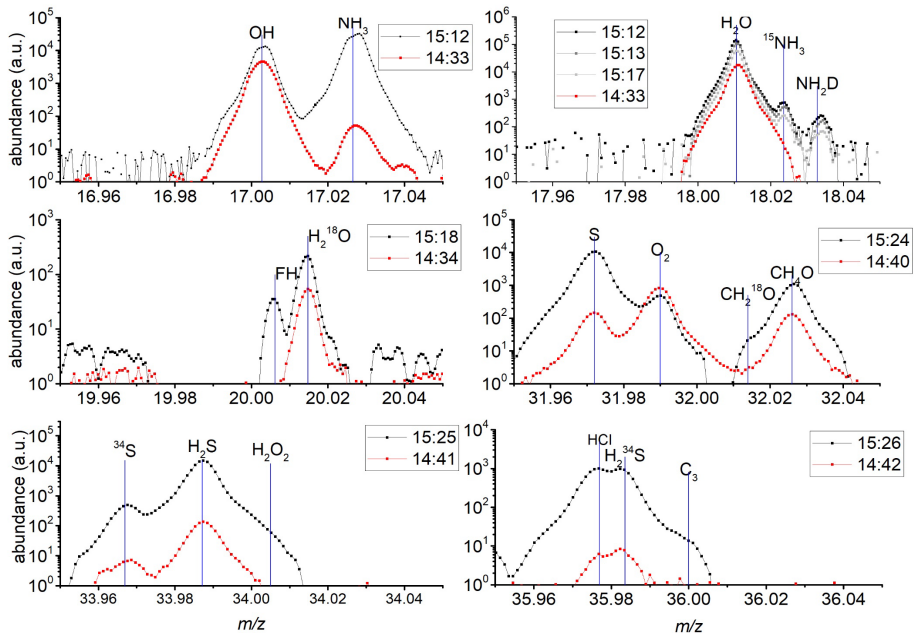

Figure 6: Mass spectra comparison after the impact (black/grey dots/lines) and before the impact ~45min earlier (red dots/lines).

Fig. 7 shows the COPS total coma density measurements from 12:30h to 18:30h on this day. The time of the impact is marked in blue. COPS showed no signs of dust (large noise peaks), although COPS is susceptible to dust (Pestoni et al., 2021a). This means that the grain hitting DFMS was not part of a big dust outburst. The overall coma density at that time was relatively low, consistent with the large distance to the nucleus (83 km), the S/C position over the northern, winter hemisphere and the relatively large distance to the Sun (2.1 au). Also plotted are the number of ions per spectrum for all *m/z* 18 measurements during this time interval. There are three *m/z* 18 spectra almost immediately following the impact, all within five minutes. The impact is clearly seen in the DFMS data, meaning that the dust grain was not devoid of water, even after many hours of being in the coma. The exponential decay time of the water signal from the dust impact was on the order of $T_{1/2}$ = 7 min. Water was back to coma background levels after ~1.5h.

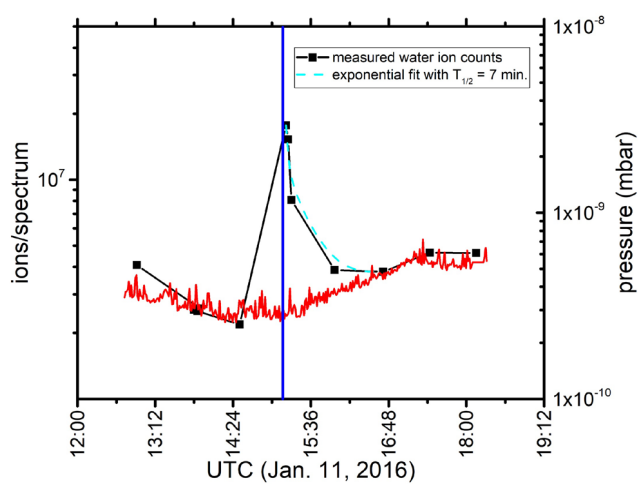

Figure 7: COPS total pressure (red) and DFMS water signal around the time of the impact (blue vertical line) on Jan. 11, 2016. The dotted, turquois line indicates an exponential trend for the water signal after the impact.

Before the impact, $H_2O$ was the most abundant molecule followed by $CO_2$, CO and $O_2$. The measured relative abundances were close to the bulk abundances derived by Rubin et al. (2019). After the impact, $H_2O$ was still the most abundant species, but closely followed by $NH_3$ and $H_2S$ (Table 2). The simultaneous increase of $NH_3$ and $H_2S$ points clearly to ammonium hydrosulphide. Looking for the ammonium salts already known from the Sep. 5, 2016 impact (Altwegg et al. 2020) we find strong increases after the impact on Jan 11, 2016 for all of them (Table 2). It is difficult to do a stoichiometric analysis, as conditions were not stable during the measurements. $NH_3$ was measured shortly after the impact, while e.g. $CH_3COOH$ was measured about 22 min. later. For all measured species except water, we have only one spectrum. We use the exponential trend line deduced for water (see Figure 7) and the change in background coma deduced

Table 2. Abundances for event # 4 in arbitrary units of relevant molecules before/after impact and their excess abundances corrected for fragmentation and time of measurement.

| m/z | Molecule | Before impact | After impact | Ratio after /before | Excess abund. corrected for fragmentation* | Normalized to $T_0$** | Coma *** | Impact material *** |
|---|---|---|---|---|---|---|---|---|
| 18.011 | $H_2O$ | 19056 | 54762 | 2.8 | 42847 | 21423 | 100 | 100 |
| 31.990 | $O_2$ | 749 | 368 | 0.5 | 0 | 0 | 3.9 | 0 |
| 17.027 | $NH_3$ | 53 | 32223 | 609 | 60157 | **30079** | 0.43 | 142 |
| 33.988 | $H_2S$ | 135 | 14700 | 109 | 27674 | 27674 | 1.1 | 129 |
| 27.011 | HCN | 92 | 2367 | 26 | 2685 | 2440 | 0.47 | 11.4 |
| 35.977 | HCl | 6 | 990 | 165 | 1525 | 1678 | 0.04 | 7.8 |
| 43.006 | CHNO | 8 | 630 | 81 | 744 | 893 | 0.04 | 4.1 |
| 46.005 | CHOOH | 2 | 200 | 100 | 772 | 927 | 0.04 | 4.3 |
| 20.006 | FH | <1 | 43 | >43 | 43 | 24 | 0 | 0.11 |
| 60.021 | $CH_3COOH$ | <1 | 14 | >14 | 51 | 77 | 0 | 0.36 |
| | *Sum of acids* | | | | | 33712 | | |
| 63.944 | $S_2$ | 5 | 270 | 54 | 265**** | 400 | 0.02 | 1.8 |
| 95.916 | $S_3$ | <1 | 31 | >31 | 31**** | 93 | 0 | 0.45 |

*Fragmentation patterns for acids from NIST (Steins et al. 2018); **$T_0$ is the time when $H_2S$ was measured; we assume a similar time dependence for all species as for water. ***abundance rel. to water in % for the coma value before impact and the excess abundance after the impact. ****Fragmentation unknown.

from COPS in order to calculate the abundances for all species at a given time. We set, arbitrarily, $T_0$

to the time when $H_2S$ was measured, which means that we have to divide e.g. $H_2O$ abundance, measured 7 minutes earlier by a factor ~2.

There are several systematic uncertainties in these observations. These uncertainties arise because most of these salts have not been calibrated with DFMS due to their poisonous and/or corrosive nature and therefore, the ionization cross sections and the fragmentation patterns are not well known. We therefore assume the same ionization cross section for all species and use the NIST tables (Steins et al. 2018) for the fragmentation patterns (see instrument section). Despite these shortcomings, the amount of excess $NH_3$ (amount of $NH_3$ from the impact minus the contribution from the undisturbed coma) matches nearly the sum of all excess acids contributing to salts, namely HF, HCN, $H_2S$, HCl, CHOOH and $CH_3COOH$. In view of the systematic uncertainties, this match is very good and we conclude, that the excess $NH_3$ after the impact is completely due to salts and that the most abundant salt is $NH_4^+SH^-$. The composition of the impact material is very different from the normal coma, with $NH_3$ being the most abundant species in the impact material followed by $H_2S$. Noteworthy is the clear absence of $O_2$ in the impact material and the large increment for $S_2$ and $S_3$.

$S_2$ in the comet seems to have two sources, a very volatile one and a more semi-volatile one as already described in Calmonte et al. (2016). $S_3$ was seen previously, but always in connection with a dusty coma during the perihelion passage. The same authors also detected $S_4$, with estimated ratios of $S_3/S_4 \approx 3$ and $S_3/S_2 \approx 3$ and of $S_3/H_2S \approx 0.01$ albeit with large uncertainties. Calmonte et al. (2016) could not distinguish between background coma and species sublimating from dust as these were not single events, but a continuous signal from the coma. Unfortunately, the mass range of DFMS was enlarged to higher masses only relatively late in the mission, so that we do not have many spectra above m/z 100 and none following a clear dust impact. That means, we could not measure $S_4$ at *m/z* 128 connected to our events.

Fluorine has been previously reported in the coma of 67P (Dhooghe et al. 2017) together with two other halogens, Cl and Br. In the outburst of Sep. 5, 2016, HCl was clearly attributed to the ammonium salt. In the current data, HCl and HF again are most likely the sublimation products of the respective ammonium salts. The acids are quite volatile, whereas the salts are much less volatile than water and exist under normal conditions (1 bar, 295K) as solids. HF is not detected in the background coma on Jan 11, 2016, but shows a clear peak after the impact (Fig. 6).

We performed similar analyses for the other three events in Table 1. The results of all 4 events are shown in Fig. 8, where we plot the ratio of the abundances of the relevant species after the event compared to suitable spectra without dust impacts. The comparison spectra were not always adjacent due to different modes of DFMS or manoeuvres of the S/C. In these cases, we looked for spectra over similar latitude/longitude as close as possible in time. In the case of the slewing, we corrected for the variations in the background coma measured with the COPS nude gauge. From Fig. 8 it is clear that in all cases ammonia and the relevant acids increased far more than water. Also plotted are the two highly volatile species $O_2$ and $CH_4$, where the former is slightly depleted during impacts, while the latter is clearly enhanced, similar to water. Although boundary conditions for the events were quite different, the results are very consistent.

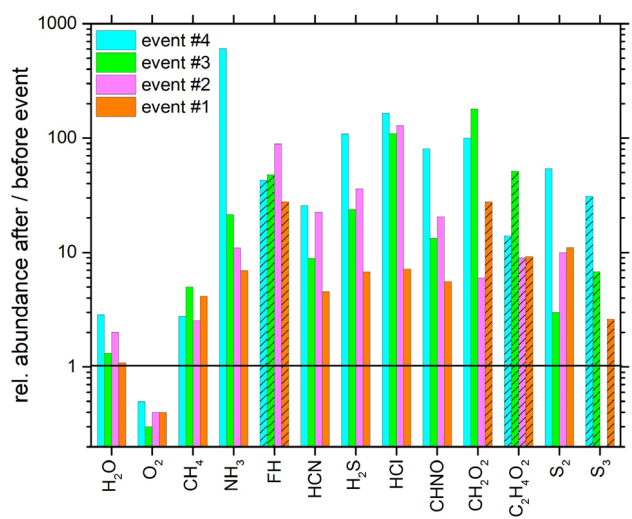

Figure 8: Comparison of abundances after the impact compared to the background coma for the 4 events. Shaded columns are lower limits as the species was not detected in the background coma.

Looking again at the balance between the abundance of $NH_3$ and acids in the impact materials, the stoichiometry is even more uncertain. In particular, we do not know when and where the grain impacted and how fast the material sublimates. We therefore ignored for the three events 1-3 any density changes in the ion source during the measurements. We still ended up with a balance between ammonia and acids well within a factor 2 (Table 3).

Other species, especially many unsaturated organics (see Fig. 6, *m/z* 98-100) are also more abundant in the impact material than in the background coma, although to a lesser degree than $NH_3$ and the acids. What is remarkable is the high abundances of organosulphur, which was already seen in the Sep. 5, 2016 outburst (Mahjoub et al. 2022). The dust grains impacting DFMS, therefore, do not consist of pure ammonium salts, but are most probably carbonaceous grains covered with ammonium salt and ice. This will be the topic of a forthcoming paper.

**Table 3**: Abundances in arbitrary units for impact events 1-3. These events could not be corrected for changes in densities during the measurements, as their time and location of impact is unknown. For event #4 see Table 2.

|  | Event# 1 | | | Event# 2 | | | Event# 3 | | |
| --- | --- | --- | --- | --- | --- | --- | --- | --- | --- |
|  | Impact[*] | Bkg[**] | Abundance corr.[***] | Impact[*] | Bkg[**] | Abundance corr.[***] | Impact[*] | Bkg[**] | Abundance corr.[***] |
| $H_2O$ | 18300 | 16903 | 2319 | 80763 | 40200 | 67335 | 12880 | 9825 | 5071 |
| $O_2$ | 636 | 1582 | 0 | 1127 | 1332 | 0 | 569 | 3432 | 0 |
| $NH_3$ | 837 | 120 | **1334** | 5386 | 350 | **9367** | 635 | 30 | **1125** |
| FH | 12 | <1 | 23 | 622 | 7 | 1156 | 16 | <1 | 30 |
| HCN | 382 | 84 | 352 | 5512 | 245 | 6215 | 237 | 27 | 248 |
| $H_2S$ | 314 | 47 | 401 | 4505 | 125 | 6570 | 863 | 36 | 1241 |
| HCl | 28 | 4 | 29 | 897 | 7 | 1068 | 73 | 1 | 86 |
| CHNO | 34 | 6 | 108 | 327 | 16 | 1200 | 71 | 5 | 255 |
| CHOOH | 12 | <1 | 12 | 120 | 20 | 100 | 120 | <1 | 120 |
| $CH_3COOH$ | 4 | <1 | 14 | 9 | <1 | 32 | 17 | <1 | 61 |
| Sum of acids |  |  | **939** |  |  | **16341** |  |  | **2041** |
| $S_2$ | 11 | <1 | 11 | 29 | 3 | 26 | 12 | 5 | 7 |
| $S_3$ | 3 | <1 | 3 | <1 | <1 | <1 | 7 | <1 | 7 |

[*]Abundance during impact, [**] Background coma abundance, [***]Excess abundance corrected for fragmentation

## 3.3 Laboratory data

We do not know the morphology of the grains impacting DFMS nor their full composition. The salts, however, are measured in the gas phase, which means that the ice sublimates from the grains most probably inside DFMS. Temperature-controlled desorption (TPD) experiments can be used to get the sublimation behaviour and the sublimation temperatures of different species, although desorption from a thin ice film is probably quite different from a complex cometary grain.

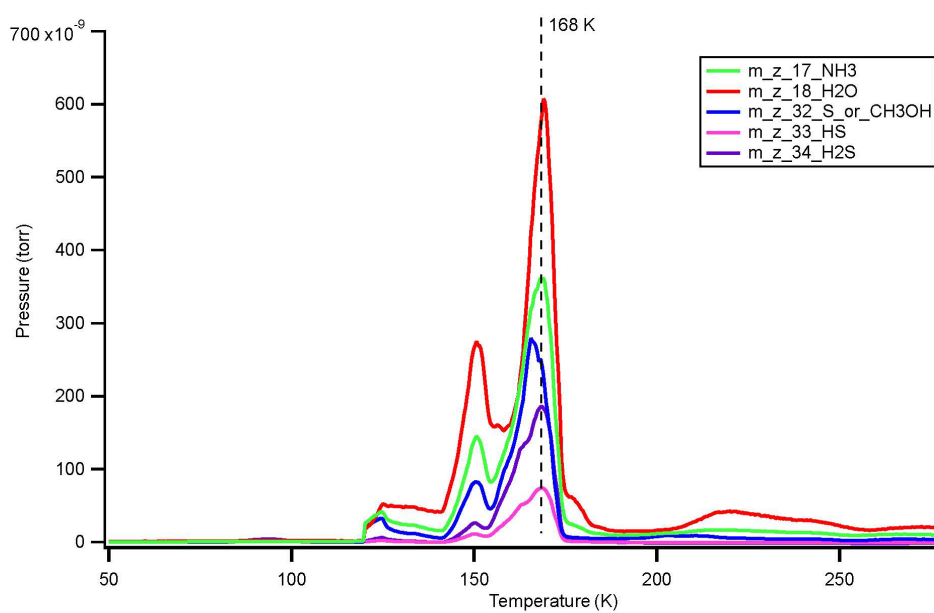

Figure 9: Temperature Programmed Desorption (TPD) curves obtained from a thin film of mixed ices $CH_3OH:H_2S:NH_3:H_2O$ (3:3:3:1) deposited at 50 K and warmed to room temperature at a rate of 0.5 K min$^{-1}$. The ion currents at *m/z* 17, 33 and 34 correspond to $NH_3$, HS and $H_2S$ respectively.

Fig. 9 shows the signature of the ammonium hydrosulphide salt in an experiment where $CH_3OH$, $NH_3$, $H_2S$ and $H_2O$ were simultaneously condensed and then released by temperature-controlled desorption (TPD). In the TPD technique, desorbing molecules from an ice film are observed as the sample temperature is increased slowly. The TPD graphs presented here are obtained from a thin film of mixed ices $CH_3OH:H_2S:NH_3:H_2O$ (3:3:3:1). The ice film was vapour-deposited on a substrate attached to the cold finger of a closed-cycle helium cryostat (ARS model DE-204). An external manifold was used to prepare gas mixtures prior to deposition. The ice film was grown by leaking the gas mixture into the chamber and forming ices on the substrate, which was held at 50 K. Immediately after deposition the sample was warmed to 300 K at a rate of 0.5 K minute$^{-1}$. A Stanford Research Systems RGA 200 quadrupole mass spectrometer was used to monitor the gaseous species released when the ice film was warmed. The mass spectrometer operated with a 70 eV electron impact ionization method scanning *m/z* 1–200. A detailed description of the experimental setup used can be found in Mahjoub et al. (2016) and Hand & Carlson (2011).

In a previous study, (Mahjoub et al. 2016) demonstrated that $NH_4^+SH^-$ is formed in the ice mixture $CH_3OH:H_2S:NH_3:H_2O$ using an ammonia isotope $^{15}NH_3$. Furthermore, (Loeffler et al. 2015) showed the formation of $NH_4^+SH^-$ salt in a $H_2S:NH_3$ ice mixture at 50 K and that this reaction is complete at 141 K. In the TPD graph, all species expected from the dissociation of $NH_4^+SH^-$ are detected at the same temperature T=168 K. These species are ammonia ($NH_3$), Hydrogen sulphide ($H_2S$) as well as fragments HS and S. The isotopologue $H_2^{34}S$ shows also a similar TPD trace (not shown here). Because the ionization cross sections and the fragmentation patterns for $NH_3$ and $H_2S$ are not equal, we cannot quantitatively compare the abundances of the two desorbing species from this plot. The peak at 150K is probably due to crystallization of water as often seen in such experiments. The signal detected at T = 168 K for both $NH_3$ and $H_2S$ is resulting from the decomposition of $NH_4^+SH^-$ salt rather than simple sublimation of pure $NH_3$ and $H_2S$ ices. Indeed, the TPD study of pure $H_2S$ and $NH_3$ ice films as reported by Collings et al. (2004) shows that both molecules are completely desorbed at T < 100 K. The formation of $NH_4^+SH^-$ stabilizes these two molecules to higher temperature around 170 K.

We conclude that the thermal degradation of $NH_4^+SH^-$ occurs at T ~ 170 K. This conclusion is in agreement with infrared measurements of $NH_4^+SH^-$ as function of temperature reported in Fig 8 of (Loeffler et al. 2015) where thermal dissociation is total around 170 K. In the same paper, TPD experiments done on irradiated ice mixtures show a slow destruction of the salt with irradiation and temperature and at the same time clear evidence for $S_3$ (Loeffler et al. 2016; Mahjoub et al. 2017).

The simultaneous desorption of $H_2S$ and $NH_3$ above the sublimation temperature of water in these experiments resembles very much what is seen during dust impacts in the coma of 67P by ROSINA/DFMS, further strengthening our findings that the signature observed in the DFMS spectra are indeed due to ammonium salts.

## 4 DISCUSSION

From our measurements, it seems that ammonia and hydrogen sulphide are embedded in cometary ice independently of each other. They show clearly different dependences on latitude and heliocentric distance. However, in cometary dust, the two molecules are very well correlated, which points to ammonium hydrosulphide salt. To determine where and when this salt formed is not straightforward. In principle, it could form at any stage prior to comet formation, during comet formation or even quite recently on the nucleus. The reaction between the acid $H_2S$ and $NH_3$ happens at almost all temperatures (e.g. Loeffler et al. 2015). It therefore is sufficient that ammonia and an acid meet each other, be it on an ice surface or on a dust grain. However, there are some indications, which may help to locate the time of formation.

a. Contemporary formation: Ammonia and acids are released from cometary ice and could settle, at least for a short time on the surface of the comet as frost, similar to e.g. $CO_2$ (Filacchione et al. 2016). On the surface, they could form salt on dust grains, which are then subsequently ejected. The lifetime of these dust grains on the comet surface is probably short, as almost all grains were detected between the two equinoxes, when the southern hemisphere was quite hot and lost up to several meters of material (Keller et al. 2017). During this time, cometary activity was high and the solar wind could not access the cometary surface. Thus, no energetic particles could impact the surface. The lack of energetic particle input makes it very unlikely that the salt in this case experienced radiolysis before being ejected. However, the abundant $S_2$ and $S_3$ in the impact material are most probably products of radiolysis of the $NH_4^+SH^-$ salt (Loeffler et al. 2015).
b. Formation in the protoplanetary disk: Before being accreted into a comet, the grains could have passed the $H_2S$ and $NH_3$ snow lines. However, in this scenario, radiation and therefore radiolysis in the midplane of the protoplanetary disk including the position where comets likely formed was quite insignificant, as Mousis et al. (2018) have modelled for the case of $H_2O$ / $O_2$. This model is most probably also applicable for the case of $NH_4SH$ / $S_n$ leaving no mechanism to form sulphur chains.

c. Prestellar origin: Dust grains could have acquired layers of salts quite early in their history. In young stellar objects, clear signs of $NH_4^+$ have been detected (Schutte & Khanna 2003). Once the salts are formed, they are much less volatile than their respective acids and ammonia (e.g. Kruczkiewicz et al. 2021). The salts may easily survive higher temperatures than required for the sublimation of water ice and other ices like e.g. $CO_2$. A prestellar origin might also be able to explain the high $S_n$ abundances as these grains probably have been exposed to radiation in the interstellar space. The water measured in the impact material could then be of a later origin, e.g. in the protoplanetary disk, where $O_2$ is not expected to exist.

The different signatures and abundances of cometary nucleus ice and cometary grain ice is intriguing, especially the missing $O_2$ in the ice on grains. While the impacts still show a considerable amount of water sublimation from grains, as well as of highly volatiles like e.g. $CH_4$ probably trapped in water (Fig. 8), even after a long travelling time in the coma, there seems to be no $O_2$ trapped in the ice of the grains. This is in contrast to the strong correlation between $O_2$ and $H_2O$ during the mission (Läuter et al. 2020; Taquet V. et al. 2016). Grains just lack the $O_2$ seen abundantly in the nucleus ice. This indicates that the ices in the nucleus and the ices on the grains formed not simultaneously and not in the same environment.

Table 4: $^{34}S/^{32}S$

| Event # | $^{34}S/^{32}S$ | Uncertainty ($\pm 1\sigma$) |
|---|---|---|
| 1 | 0.0446 | 0.0093 |
| 2 | 0.0507 | 0.0076 |
| 3 | 0.0472 | 0.0080 |
| 4 | 0.0484 | 0.0033 |

The dust impact events are dynamic and therefore not ideal to determine isotopic ratios for species, which are not measured simultaneously, even if the time difference is only small (<1 min.). Because the sulphur signal is quite high, we nevertheless derived $^{34}S/^{32}S$ for all 4 events, albeit with large uncertainties. The results are given in Table 4. The uncertainties contain statistical uncertainties and fit uncertainties. They do not contain systematic uncertainties due to the changing density with time. All values are higher than the mean value of 0.0423±0.007 derived for the average of pure volatiles in Calmonte et al. (2017), although #1 and #3 overlap within 1σ due to large error bars. They, however, fall all within the range 0.0463 ± 0.0057 as determined by the COSIMA instrument for four cometary dust particles in 67P (Paquette et al. 2017). This might be another indication - to be taken with caution - that the ice on the grains did not form in the same environment as the nucleus ice.

Nitrogen abundance in comets has been found to be depleted relative to the solar N/O ratio. However, the $NH_3/H_2O$ abundance in the coma increases with decreasing heliocentric distance (Altwegg et al. 2020). One explanation could indeed be a reservoir of nitrogen in the form of ammonium salts, which are generally less volatile than $NH_3$ and their respective acids. This would mean that $NH_3$ covering dust grains is locked in salt except for small heliocentric distances. Sulphur also remains a mystery. In star forming regions, sulphur seems to be depleted relative to cosmic abundances by several orders of magnitude. Considering the coma of 67P and taking into account the fraction of sulphur in the refractories, sulphur seems to be close to cosmic abundance (Calmonte et al. 2016). The uncertainty is, however, high as the amount of sulphur in refractories for 67P is unknown and was derived from the flyby at comet 1P/Halley. The presence of abundant $NH_4^+SH^-$ may solve this riddle, as this salt on dust grains is hard to detect remotely in star forming regions.

## 5 CONCLUSIONS

Sudden outbursts of $NH_3$ simultaneously with $H_2S$ detected with the ROSINA-DFMS instrument on the Rosetta S/C point to the presence of abundant ammonium hydrosulphide in or on carbonaceous grains from comet 67P/Churyumov-Gerasimenko. There seems to be a clear distinction between the nucleus ice, where $H_2S$ and $NH_3$ exist independently and grains, where they desorb together. $S_2$ is much more abundant on grains compared to water than in the ice of the comet, while $S_3$ is found only in grain impacts. This higher abundance points to radiolysis in these grains, which means they must have been exposed to energetic particles over an extended time. While for operational reasons, $S_4$ could not be measured close to the dust impacts, $S_4$ was clearly identified in periods where the coma was very dusty (Calmonte et al. 2016). Longer sulphur chains very likely are refractory, not sublimating at temperatures reached in the instrument or on grains in the coma. While $S_n$ can also be formed from pure $H_2S$ ice by photo processing (Cazaux et al. 2021), the fact that $S_3$ is clearly related to dust and is not found in the normal nucleus ice, where $H_2S$ is quite abundant, indicates that $S_3$ is a product of radiolysis of the ammonium salt. In addition, photo processing of $H_2S$ results not only in $S_n$, but also in $H_2S_2$ (Cazaux et al. 2021), a species not detected in the DFMS *m/z* 66 and *m/z* 65 ($HS_2$) spectra. This exposure rules out a contemporary formation of the salt on the surface or in the interior of the comet or a formation of the salt in the midplane of the protoplanetary disk, while the comet accreted. A prestellar formation is therefore likely. The salt is semi-volatile, less volatile than water and could probably have survived quite high temperatures. It seems that on these grains, acids and ammonia are all locked in salts, be it sulphur, halogens or carbon bearing acids like HOCN. If indeed, a relatively large part of sulphur and nitrogen is therefore in a semi-volatile state in these grains, then the depletion of nitrogen in comets and of sulphur in star forming regions could probably be explained, primarily because salts escape detection unless they experience temperatures above water sublimation. With the JWST S/C in orbit, there is hopefully the possibility to detect salts, or at least several of the acids in ices, which are supposed to be part of ammonium salt, like HOCN, $H_2CO$ and formamide while looking for ammonium salts in star forming regions and possibly comets.

**Acknowledgements.**

Work at the University of Bern was funded by the State of Bern and the Swiss National Science Foundation (200020_207312). S.F.W. acknowledges the financial support of the SNSF Eccellenza Professorial Fellowship (PCEFP2_181150). Part of this research was carried out at the Jet Propulsion Laboratory, California Institute of Technology, under a contract with the National Aeronautics and Space Administration (80NM0018D0004). A.M. is thankful for NASA DDAP financial support. This paper is a homage to the first PI of ROSINA, Hans Balsiger, whom we sadly lost in 2021. The results from ROSINA would not be possible without the work of the many engineers, technicians, and scientists involved in the mission, in the Rosetta spacecraft, and in the ROSINA instrument team over the past 20 years, whose contributions are gratefully acknowledged.  Rosetta is a European Space Agency (ESA) mission with contributions from its member states and NASA. We thank herewith the work of the whole ESA Rosetta team.

**Data availability:**

All ROSINA flight data have been released to the Planetary Science Archive of ESA (PSA) and to the Planetary Data System (PDS) archive of NASA. The data used in this article are available in the European Space Agency's PSA at archives.esac.esa.int/psa/.

**References**

Altwegg K. et al., 2020, Nature Astronomy, 4(5), 533-540

Atreya S.K. et al., 1999, Planet. Space Sci. 47, 1243–1262

Balsiger H. et al., 2007, SSR, 128(1), 745-801

Berg B. L. et al., 2016, Icarus, 265, 218-237

Calmonte U. et al., 2016, MNRAS, 462(Suppl_1), S253-S273

Calmonte U. et al., 2017, MNRAS, *469*(Suppl_2), S787-S803.

Cazaux S. et al., 2021, arXiv preprint arXiv:2110.04230

Clementi E. & Gayles J. N., 1967, J. Chem. Phys., 47, 3837

Collings M. P. et al., 2004, MNRAS, 354, 1133

De Keyser J. et al., 2019, Int. J. Mass Spectrom., 446, 116232

Della Corte V. et al., 2016. MNRAS, 462(Suppl_1), S210-S219

Dhooghe F. et al., 2017, MNRAS, 472(2), 1336-1345

Filacchione G. et al., 2016, Science, 354(6319), 1563-1566

Hand K. and Carlson R.W., 2011, Icarus, 215, 226

Hänni N. et al., 2019, The Journal of Physical Chemistry A, 123(27), 5805-5814

Keller H. U. et al., 2017, MNRAS, 469(Suppl_2), S357-S371

King T. V., Clark R. N., Calvin W. M., Sherman D. M., & Brown R. H., 1992, Science, *255*(5051), 1551-1553

Kruczkiewicz F., Vitorino J., Congiu E., Theulé P., Dulieu F. 2021, A&A, 652, A29

Läuter M., Kramer T., Rubin M., & Altwegg K., 2020, MNRAS, 498(3), 3995-4004

Le Roy L. et al., 2015, A&A, 583, A1

Loeffler M. J., Hudson R. L., Chanover N. J., Simon A. A., 2015, Icarus, 258, 181-191

Loeffler M. J., Hudson R. L., Chanover N. J., Simon A. A.,2016, Icarus, 271, 265-268

Mahjoub A. et al. 2016, ApJ 820, 141

Mahjoub A. et al., 2017, ApJ, 846, 148, http://dx.doi.org/10.3847/1538-4357/aa85e0

Mahjoub A. et al., 2022, submitted

Minissale M. et al., 2022, ACS Earth and Space Chemistry 6 (3), 597-630

Mousis O. et al., 2018, ApJ 858(1), 66

Paquette, J. A.et al., 2017, MNRAS, *469*(Suppl_2), S230-S237.

Pendleton Y. J., Tielens A. G. G. M., Tokunaga A. T., Bernstein M. P., 1999, ApJ, 513(1), 294

Pestoni B. et al., 2021a, A&A, 645, A38

Pestoni B. et al., 2021b, A&A, 651, A26

Poch O. et al., 2020, Science, 367(6483), eaaw7462

Roman M.T., Banfield D., Gierasch P.J., 2013, Icarus 225, 93–110

Rubin M. et al, 2019, MNRAS, 489(1), 594-607

Schutte W. A. & Khanna R. K., 2003, A&A, *398*(3), 1049-1062

Steins S. E., 2018, NIST Standard Reference Database Number 69, National Institute of Standards and Technology

Weidenschilling S.J., Lewis J.S., 1973, Icarus 20, 465–476